\documentclass[%
 reprint,
 superscriptaddress,
 amsmath,amssymb,
 aps,
 prb,
]{revtex4-2}

\usepackage{graphicx}
\usepackage{dcolumn}
\usepackage{bm}
\usepackage{dsfont}
\usepackage{braket}
\usepackage{hyperref}
\usepackage{xcolor}
\usepackage{soul}
\usepackage{changes}
\usepackage{ulem}
\usepackage{dsfont}

\newcommand{\RNum}[1]{\uppercase\expandafter{\romannumeral #1\relax}}
\let\vec\mathbf

\begin{document}

\title{Control of Plasmons in Doped Topological Insulators via  Basis Atoms}

\author{Zhihao Jiang}
\affiliation{Department of Physics and Astronomy, University of Southern California, Los Angeles, California 90089-0484, USA}
\author{Henning Schl\"{o}mer}
\affiliation{Institute for Theoretical Solid State Physics, RWTH Aachen University, 52056 Aachen, Germany}
\author{Stephan Haas}
\affiliation{Department of Physics and Astronomy, University of Southern California, Los Angeles, California 90089-0484, USA}

\date{\today}

\begin{abstract}
Collective excitations in topologically non-trivial systems have attracted considerable attention in recent years. Here we study plasmons in the Su-Schrieffer-Heeger model whose low-energy electronic band is only partially filled, such that the system is metallic. Using the random phase approximation, we calculate the intra- and inter-band polarization functions and determine the bulk plasmonic dispersion from the dielectric function within the random phase approximation. We find that the sub-lattice basis states strongly affect the polarization functions and therefore control the system's plasmonic excitations. By varying the real-space separation of these local orbitals, one can thus selectively enhance or suppress the plasmonic energies via a tunable trade-off between intra-band and inter-band screening processes. Specifically, this mechanism can be used to stabilize undamped high energy plasmons that have already been reported in related models. We propose scenarios on how to control and observe these effects in experiments.      
\end{abstract}

\maketitle

\section{\label{sec-intro} Introduction}
Topological insulators (TIs) have been studied extensively  in the last few decades, starting with the discovery of the integer quantum Hall effect \cite{Klitzing1980, Tsui1982, Kane2005, Kane2005qshe_graphene, Zhang2005, Bernevig2006, Bernevig2006HgTe, Konig2007, Novoselov2007, Fu2007, Fu2007TI_IS, Hsieh2008, Chen2009_3DTI, Kuroda2010, Gong2019, Noguchi2019, Hasan2010RMP, Qi2011RMP}. They are characterized by fundamentally non-trivial electronic states, which can be detected via bulk topological invariants and corresponding symmetry protected edge states. Recently, there has been increased interest in the collective excitations of topological systems, such as plasmons \cite{Karch2011, Okada2013, Schutky2013, Efimkin2012, Yuan2017EXP, Dubrovkin2017EXP, Jin2017, DiPietro2013EXP, Venuthurumilli2019EXP, Jiang2020, Schlomer2020plasmons, Lai2014, Ginley2018EXP, Qi2014}, excitons \cite{Klembt2018, Knolle2017, Kartashov2017, Kung2019}, and magnons \cite{Ruckriegel2018, Malki2019, Mook2014}. Propagating plasmons on the surfaces of two-dimensional (2D) and three-dimensional (3D) TIs \cite{Karch2011,Schutky2013,Jin2017,Efimkin2012,Qi2014}, as well as localized plasmons on the ends of one-dimensional (1D) TIs \cite{Jiang2020}, have been theoretically investigated and partly observed in experiments \cite{Yuan2017EXP,Ginley2018EXP,DiPietro2013EXP,DiPietro2013EXP,Dubrovkin2017EXP}. These plasmons have been shown to inherit some important topological features from the host materials \cite{Jin2017,Jiang2020}, for example, chiral one-way propagation without back-scattering of the surface modes and robustness of localized modes against disorder. These findings indicate that the non-trivial topology of the single-electron states have a clear impact on their collective excitations on the surface of the materials.

In this work, we study the plasmonic excitations in doped topological insulators, focusing on their \textit{bulk dispersion} in  momentum space. Within the tight-binding approach, an insulator is described as a multi-band model whose external periodic structure defines the reciprocal lattice in momentum space, and whose internal basis orbitals determine the properties of the electronic eigenstates for each lattice momentum $k$ \cite{Cohen2016Book}. The latter feature will be the main focus of this paper because it is non-trivial in TIs. The modulation of the eigenstates in $k$-space determines the topological invariants, for example, the Chern number of 2D quantum Hall insulators and the winding number of 1D TIs \cite{Asboth2016Book}. These eigenstates also enter the calculation of the charge-charge correlation function, thus ultimately affecting the dielectric response function and plasmonic excitations in the material. The main motivation of this paper is to examine this dependence of plasmons on the internal orbitals. For the sake of simplicity, we choose the Su-Schrieffer-Heeger (SSH) model \cite{Su1979}, which is a 1D bi-particle chain with alternating hopping parameters. The model has two bands due to its two internal orbitals, namely, two basis atoms within each primitive cell. All the eigenenergies and eigenstates can be derived analytically. By doping, we partially fill the low-energy band, such that the system is in a metallic regime. This opens both the intra- and inter-band polarization channels, which will be separately analyzed in detail. We then tune the internal orbitals by changing their real-space separation within the primitive cell while keeping the hopping amplitudes constant, which has a remarkable impact on the plasmonic excitations. Despite the simplicity of the model and calculations, the underlying ideas are general and are expected to be applicable to more complex   situations.

The paper is organized as follows. In the next section, we discuss a general method, based on the random phase approximation, of calculating plasmonic excitations in multi-band tight-binding models. We focus on the main ingredient, i.e., the \textit{overlap function}, which captures the information of internal orbitals and directly affects the plasmon calculations. Following that, we consider specifically the SSH model in Sec.~\ref{sec-3Results}. The overlap functions and electron energy loss spectra (EELS) are calculated analytically and numerically. We examine in detail how plasmon dispersions are affected by varying the inter-orbital separation. In Sec.~\ref{sec-4Conclu}, we give  conclusions and discuss possible generalizations as well as future directions.

\section{\label{sec-2Model} General Method}
\begin{figure}[htbp]
 \includegraphics[width=8.6cm]{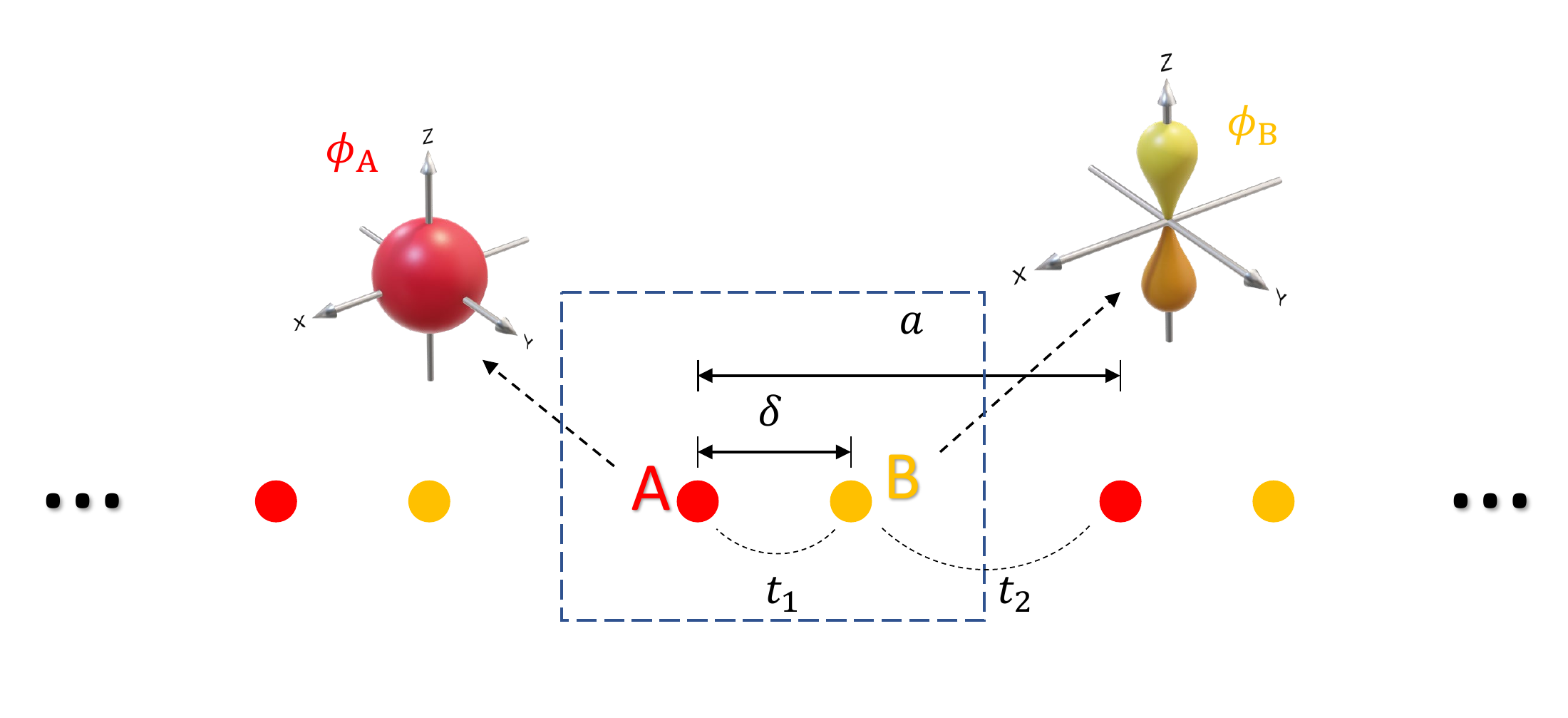}
 \caption{Illustration of a one-dimensional lattice with two distinct atoms A and B in a primitive cell.}\label{fig1_sketch}
\end{figure}

To introduce the technical approach, let us consider a  one-dimensional (1D) lattice with a basis, whose primitive cell consists of two distinct atoms, A and B, which can  host different orbitals $\phi_\text{A}$ and $\phi_\text{B}$, as illustrated in Fig.~\ref{fig1_sketch}. The  method can be generalized to any dimensionality with more than two basis orbitals in a primitive cell. We set the lattice spacing to be $a$ and the separation between two atoms in the primitive cell to be $\delta$. This simple structure can be parameterized to describe some well-known models, such as the SSH model~\cite{Su1979} and the Rice-Mele model~\cite{Rice1982}. To solve the electronic eigenenergies and wavefunctions, we follow the standard tight-binding approach by forming a linear combination of the atomic orbitals (LCAO) for each atom basis~\cite{Cohen2016Book},
\begin{align}
    \psi_{\vec k}^{\text{A/B}}(\vec r) = \frac{1}{\sqrt{N}} \sum_n  e^{i{\vec k \cdot \vec R_{n}^{\text{A/B}}}} \phi_{\text{A/B}}(\vec r - \vec R_{n}^{\text{A/B}} ),
\end{align}
where $\vec R_{n}^{\alpha}$ is the position of atom $\alpha$ ($\alpha \in \{\text{A, B}\}$) in the $n$-th unit cell, and $\vec k$ is the momentum vector in the first Brillouin zone (1BZ). From Fig.~\ref{fig1_sketch} it is clear that $\vec R_{n}^{\text{B}} - \vec R_{n}^{\text{A}} = \bm{\delta} = \delta \hat{x}$. These LCAOs serve as basis functions in which we can expand the electronic states and write down the Hamiltonian matrix. For each momentum $\mathbf{k}$, the Hilbert space is spanned by the basis functions $\psi_{\vec k}^{\text{A}}(\vec r)$ and $\psi_{\vec k}^{\text{B}}(\vec r)$. Hence, the Hamiltonian with nearest neighbor hoppings is given by
\begin{align}
    H(\vec k) = 
    \begin{pmatrix}
               V_\text{A} & t_1 e^{i \vec k \cdot \bm{\delta}} + t_2 e^{-i\vec k \cdot (\vec a -\bm{\delta})} \\
               t_1^* e^{-i \vec k \cdot \bm{\delta}} + t_2^* e^{i\vec k \cdot (\vec a - \bm{\delta)}} & V_{\text{B}}
    \end{pmatrix} . \label{HamiltonianHk}
\end{align}
In principle, $V_\text{A}$, $V_\text{B}$, $t_1$ and $t_2$ can be obtained from first principle calculations or determined from experiments. Here, we treat them as tunable parameters which can be chosen at will to analyze plasmonic properties in various parameter regimes. Later on, we will discuss the connection between these results to realistic experiments. 

Diagonalizing $H(\vec k)$ yields the electronic eigenenergies $E_n(\vec k)$ and eigenstates $\Phi_n(\vec k)$. We can further evaluate the free-electron polarization (matrix) function $\bm{\chi}_0(q,\omega)$, whose elements in the sub-lattice basis are explicitly given by
\begin{eqnarray}
 [\bm{\chi}_0(q,\omega)]_{\alpha \beta} & = & 2\sum_{k,n,n'} \frac{f_{n',k+q}-f_{n,k}}{E_{n',k+q}-E_{n,k}-\omega-i0^+} \nonumber \\
 &\times &\Phi_{n,k,\alpha}^* \Phi_{n',k+q,\alpha} \Phi_{n',k+q,\beta}^* \Phi_{n,k,\beta}. \label{Pi0_q}
\end{eqnarray}
The factor of 2 arises from the spin degeneracy, and $f_{n,k}$ is the Fermi function. $\alpha, \beta \in \{\text{A, B}\}$ are sub-lattice indices, which are retained here because the Fourier transformation is only performed over the periodic part of the lattice defined by $\vec R_{n}$.

We can then write the Coulomb interaction $\bm{V}(q)$ in the same basis, which, in principle, is also a $2 \times 2$ matrix with elements $V_{\alpha \beta}$. However, in the basis constructed from the LCAOs, all the Coulomb elements are identical and do not depend on the sub-lattice indices in the continuum limit, namely $\bm{V}_{\alpha \beta}(q)\equiv V(q)$. This allows us to finally arrive at a scalar dielectric function,
\begin{align}
\varepsilon(q,\omega) = 1 - V(q) \chi_0(q,\omega), \label{Eps_RPA_q} 
\end{align}
within the random phase approximation (RPA). Here, $\chi_0(q,\omega)$ is the bare polarization function including all components between pairs of sub-lattice indices. It is given by
\begin{align}
 \chi_0(q,\omega) = 2\sum_{k,n,n'} \frac{(f_{n',k+q}-f_{n,k})O_{nn'}(k,k+q)}{E_{n',k+q}-E_{n,k}-\omega-i0^+} \label{Chiq},
\end{align}
where we define the overlap function (sometimes also called band coherence factor~\cite{Lewandowski2019}),
\begin{align}
    O_{nn'}(k,k+q) \equiv |\braket{\Phi_{n,k}|\Phi_{n',k+q}}|^2. \label{ovpGeneral}
\end{align}
Plasmons are defined as singularities of the electron energy loss spectrum (EELS),  
\begin{align}
  \text{EELS}(q,\omega) = -\mathrm{Im} \frac{1}{\varepsilon(q,\omega)}.  
\end{align}
Note that the overlap functions in Eq.~\eqref{ovpGeneral} fully capture the information of internal orbitals, such as the separation parameter $\delta$, which is why it is the central observable discussed below.

\section{\label{sec-3Results}Results and Discussion}
\subsection{Overlap Functions in the SSH Model}
\begin{figure}[htbp]
 \includegraphics[width=8.6cm]{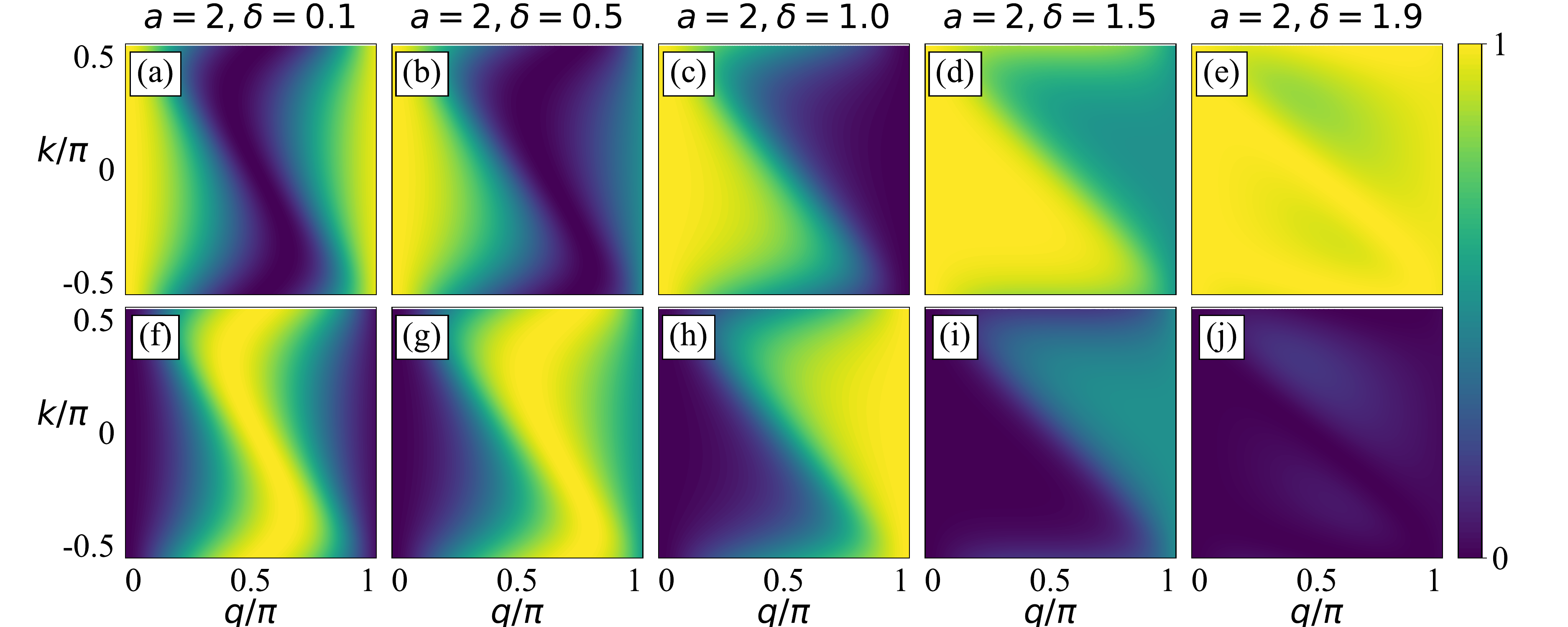} 
 \caption{Overlap functions in the SSH model for different  basis atom separations, $\delta$, with hoppings $t_1$ and $t_2$ set to be $0.5\ \text{eV}$ and $1.5\ \text{eV}$. (a-e) Intra-band overlap functions. (f-j) Inter-band overlap functions.}\label{fig2_ovpSSH}
\end{figure}
Let us now consider the Su-Schrieffer-Heeger (SSH) chain with unequal hopping matrix elements, $t_1 \neq t_2$, while $V_\text{A}=V_\text{B}=0$. The bulk model has two electronic bands separated by an energy gap, $2|t_1- t_2|$. We denote the bonding low energy band as $E_{0,k}$ and the anti-bonding high energy band as $E_{1,k}$. They are related via $E_{0,k}=-E_{1,k}$ due to the chiral symmetry \cite{Asboth2016Book}. For this system, the overlap function has the form
\begin{align}
  O_{nn'}(k,k+q) = 
  \begin{cases}
    \frac{1}{2}  + \frac{1}{2} \cos(\varphi_{k,q})  & \quad \text{if } n=n', \\
    \frac{1}{2}  - \frac{1}{2} \cos(\varphi_{k,q})  & \quad \text{if } n \neq n',
  \end{cases} \label{ovpSSH}
\end{align}
where
\begin{align}
    \varphi_{k,q} &= q\delta + \mathrm{arg}(z_{k,q}) \nonumber \\
    &= q\delta + \mathrm{arg}\bigg (\frac{t_1 + t_2e^{-ika-iqa}}{t_1 + t_2e^{-ika}} \bigg ). \label{relative_phase}
\end{align}
An open-ended SSH chain has two topologically distinct sectors corresponding to $t_1 > t_2$ and $t_1 < t_2$, i.e.,  a topologically trivial sector with no localized edge states, and a non-trivial sector with two edge states. The overlap functions closely relate to the topology of the system through the second term in Eq.~\eqref{relative_phase}. This can be easily understood in the two limiting cases: (1) when $t_1/t_2 \to 0$, $\mathrm{arg}(z_{k,q}) \to -qa$; and (2) when $t_2/t_1 \to 0$, $\mathrm{arg}(z_{k,q}) \to 0$. Furthermore, the overlap function  depends on the real space configuration of the model, i.e., $\delta$, via the first term in Eq. \eqref{relative_phase}. 

In Fig.~\ref{fig2_ovpSSH}, we show the overlap functions $O_{nn'}(k,k+q)$ for both intra-band [$n=n'$, Figs.~\ref{fig2_ovpSSH}(a)-(e)] as well as inter-band transitions [$n\neq n'$, Figs.~\ref{fig2_ovpSSH}(f)-(j)], while scanning through different  intra-cell atom separations $\delta$, i.e., tuning the distance between the basis atoms. Here, the hopping parameters are set to  $t_1=0.5\ \text{eV}<t_2=1.5\ \text{eV}$. It can be clearly seen how the intra-cell separation $\delta$ affects the overlap functions. For large $\delta$, the intra-band overlap is close to $1$ [Fig.\ref{fig2_ovpSSH}(e)] over the entire $(k,q)$ parameter space, whereas inter-band correlations are close to $0$ [Fig.\ref{fig2_ovpSSH}(j)]. From  Eq.~\eqref{ovpSSH}, it follows that there is a complementarity of the intra- and inter-band overlap functions in the SSH model, namely, they add up to 1. Therefore, when the intra-band (inter-band) correlations are enhanced, the inter-band (intra-band) overlaps are suppressed and vice versa. This complementarity is  generally true for multi-band models in any dimensionality, which can be proved by using Eq.~\eqref{ovpGeneral} and the completeness of electronic eigenstates, i.e. for any band $n$,
\begin{align}
    \sum_{n'}O_{nn'}(k,k+q) 
= \sum_{n'} |\braket{\Phi_{n,k}|\Phi_{n',k+q}}|^2 
    = 1.
\end{align}
Below, it will be shown  that the strongly dimerized (large $\delta$) regime supports prominent low-energy plasmons, whereas high-energy plasmons are suppressed due to small inter-band overlaps. For small $\delta$, the intra-band overlap function is globally decreased, especially for intermediate $q$ [Fig.\ref{fig2_ovpSSH}(a)]. Inter-band correlations, on the other band, increase [Fig.\ref{fig2_ovpSSH}(f)]. This implies that, for small $\delta$, the low-energy plasmons experience stronger inter-band screening and have a softened dispersion. The high-energy plasmons, however, become prominent in this regime. 

As switching $t_1$ and $t_2$ is equivalent to a mapping  $\delta \to a-\delta$ in a periodic system, we do not need to consider $t_1 > t_2$ as a separate case. In other words, the overlap functions for $t_1 > t_2$ show the same behavior as in Fig.~\ref{fig2_ovpSSH} for $t_1 < t_2$, with $\delta$ replaced by $a-\delta$. Specifically, when $\delta=1=a/2$, switching $t_1$ and $t_2$ has no effect on the bulk properties of the model. However, when the model is cut into a finite chain with open boundaries, switching $t_1$ and $t_2$ fundamentally changes the electronic as well as the plasmonic properties on the system \textit{boundaries} \cite{Asboth2016Book,Jiang2020}. 

\subsection{Plasmonic Excitations in the SSH Model}
\begin{figure}[htbp]
 \includegraphics[width=8.6cm]{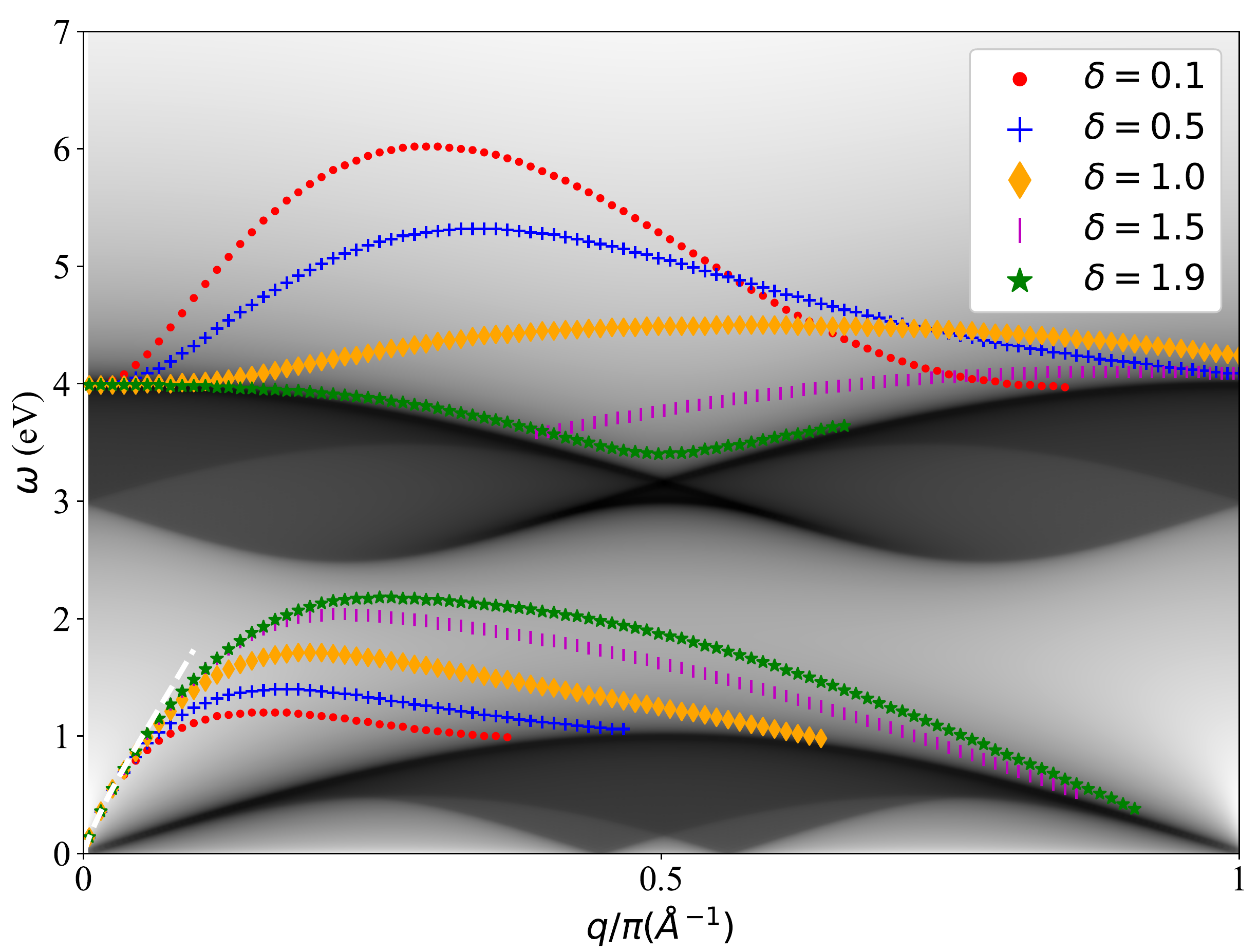} 
 \caption{Plasmon dispersions in the SSH chain for different basis atom separations, $\delta$, with $t_1$ and $t_2$ set to be $0.5$eV and $1.5$eV (same parameters as in Fig. 2). The dark areas represent the electron-hole continua.}\label{fig3_eelsSSH}
\end{figure}
We now discuss how the overlap functions affect the plasmonic excitations in the SSH model. The dielectric function can be evaluated using Eqs.~\eqref{Eps_RPA_q}, ~\eqref{Chiq} and ~\eqref{ovpSSH}. The Fermi level is set to  $E_F=-1.5\ \text{eV}$, such that the low-energy band $E_{0,k}$ is partially filled while the high-energy band $E_{1,k}$ is empty. In this case, we have both intra-band and inter-band contributions to the polarization function, which will be analyzed separately below. 

We first examine in detail the low-energy plasmons which arise from the intra-band polarization within the metallic band $E_{0,k}$, subject to the inter-band dielectric screening. Too see this, we write the full dielectric function as
\begin{align}
 \varepsilon(q,\omega) &= 1 - V(q)[\chi^\text{intra}_0(q,\omega)+\chi^\text{inter}_0(q,\omega)] \nonumber \\
 &= \varepsilon^\text{inter}(q,\omega)[1-W(q,\omega)\chi^\text{intra}_0(q,\omega)],
\end{align}
where $\varepsilon^\text{inter}(q,\omega)=1-V(q)\chi^\text{inter}_0(q,\omega)$ is the inter-band dielectric function, and $W(q,\omega)=[\varepsilon^\text{inter}(q,\omega)]^{-1}V(q)$ is the screened Coulomb interaction. For the bare Coulomb interaction $V(q)$ in  1D momentum space, we use $V(q)=2e^2K_0(ql)/a$~\cite{Santoyo1993}, where $K_0(\cdot)$ is the zeroth modified Bessel function of the second kind, which describes a quasi-1D model with a cross-section $l^2$ and $l \ll chain\ length$. In our simulations, $l$ is taken to be half of the lattice spacing $a$. For low-energy plasmons, we can approximately ignore the frequency dependence in $\varepsilon^\text{inter}(q,\omega)$ and just keep $W(q)\approx W(q,\omega=0)$, i.e., the statically screened Coulomb potential. The inter-band dielectric function then takes the simplified form
\begin{align}
  \varepsilon^{\text{inter}}(q) \simeq 1 - 4V(q)\sum_{k=-k_F}^{k_F} \frac{O_{01}(k,k+q)}{E_{0,k}+E_{0,k+q}}, \label{ssh_inter}
\end{align}
where
$O_{01}(k,q)$ is the inter-band overlap function which can be evaluated from Eq.~\eqref{ovpSSH} and which is plotted in Figs.~\ref{fig2_ovpSSH}(f)-(j). In deriving Eq.~\eqref{ssh_inter}, we made use of the chiral symmetry of the energy bands, $E_{0,k}=-E_{1,k}$. 

The intra-band polarization can be evaluated by setting $n=n'=0$ in Eq.~\eqref{Chiq}, resulting in
\begin{align}
    \chi^{\text{intra}}_0(q,\omega) \simeq \sum_{k=-k_F}^{k_F} \frac{4(E_{0,k+q}-E_{0,k})O_{00}(k,k+q)}{(\omega+i0^+)^2-(E_{0,k+q}-E_{0,k})^2}. \label{ssh_intra}
\end{align}
Here, $O_{00}(k,k+q)$ is the intra-band overlap function, illustrated in Figs.~\ref{ovpSSH}(a)-(e). The plasmon frequencies $\omega_p$ then solve the equation 
\begin{align}
    \mathrm{Re}[\varepsilon(q,\omega)] \propto 1 - W(q)\mathrm{Re}[\chi^{\text{intra}}_0(q,\omega)] = 0. \label{ReEPS_ssh}
\end{align}

In Fig.~\ref{fig3_eelsSSH}, we show numerical results for the plasmon dispersions of the SSH model for different basis atom separations, $\delta$. Here, we observe that changing $\delta$ affects both the low-energy and the high-energy plasmons. Generally, as one branch becomes more prominent, the other branch is suppressed, which can be attributed to the complementarity of the intra- and inter-band overlap functions. For larger $\delta$, the low-energy plasmon branch lies further outside the intra-band electron-hole continuum and therefore remains undamped for a larger range of momenta $q$. In this regime, we can see that the intra-band overlap function is large over the entire $(k,q)$ space, whereas the inter-band overlap function is almost zero globally [see Figs.~\ref{ovpSSH}(e) and (j)]. This results in negligible inter-band screening effects on the low-energy plasmons in this regime. In contrast, the high-energy plasmon branch becomes more prominent with gradually decreasing $\delta$. And meanwhile, the low-energy plasmon branch is softened and moves closer to the electron-hole continuum. This is due to the enhanced inter-band overlap function and reduced intra-band overlap function [see Figs.~\ref{ovpSSH}(a) and (f)]. Thus, we can tune the parameter $\delta$ to selectively enhance the low-energy or the high-energy plasmon branch and suppress the other one. 

Next, we give a semi-analytical derivation of the low-energy plasmon frequencies, in which the effects of the overlap functions show up explicitly. First, as we are interested in the plasmon solutions $|\omega_p|^2\gg(E_{0,k+q}-E_{0,k})^2$ \cite{DaJornada2020}, where we simplify the real part of the intra-band polarization function to
\begin{align}
   \mathrm{Re}[\chi^\text{intra}_0(q,\omega_p)] \approx \frac{4}{\omega_p^2}\int_{k_F-q}^{k_F} &dk\ g(k)O_{00}(k,k+q) \nonumber \\
   &\times (E_{0,k+q}-E_{0,k}).
\end{align}
Here, $g(k)$ is the density of states. For the half-filled low energy band, we make a linear approximation to the energy difference around $k_F$,
\begin{align}
    E_{0,k+q}-E_{0,k} \approx \partial_k E_{0,k}|_{k_F}q = v_Fq,
\end{align}
which is adequate for an appreciable range of $q$ ($|q|\lessapprox 0.4\ \AA^{-1}$). This leads to 
\begin{align}
   \mathrm{Re}[\chi^\text{intra}_0(q,\omega_p)] &\approx \frac{4q v_F}{\omega_p^2}\int_{k_F-q}^{k_F} dk\ g(k)O_{00}(k,k+q). \label{chi_approx}
\end{align}
Putting this back into Eq.~\eqref{ReEPS_ssh}, we find that the plasmon frequencies are approximately given by
\begin{align}
    \omega_p^2 \approx 4qW(q)v_F \int_{k_F-q}^{k_F} dk\ g(k)O_{00}(k,k+q).
\end{align}

From this expression, we recognize an explicit dependence of $\omega_p$ on the Fermi velocity $v_F$, the screened Coulomb interaction $W(q)$ which is negatively correlated with the inter-band overlap function, and an integral of the intra-band overlap function over a range $q$ around $k_F$. Thus, the overlap functions affect plasmons through at least two aspects. First, the screened Coulomb interaction $W(q)$ depends on the inter-band overlap function. When the inter-band overlap function is large, the screening effect is stronger and the Coulomb interaction $W(q)$ is further reduced, which in turn leads to a softening of the plasmon dispersion. This corresponds to the small $\delta$ limit for intermediate $q$ [for example, see Figs.~\ref{ovpSSH}(f) and (g)]. In contrast, for large $\delta$, the inter-band screening effect is weak due to the reduced inter-band correlations [Figs.~\ref{ovpSSH}(i) and (j)]. Second, the plasmon frequency is proportional to the integral of the intra-band overlap function, $\int_{k_F-q}^{k_F} dk\ g(k)O_{00}(k,k+q)$, around the Fermi vector $k_F$. Thus, in general, for larger intra-band overlaps, the plasmon dispersion is hardened to higher energies above the electron-hole continuum.  

The dependence of plasmon excitations on the overlap functions may not be easily noticeable in ordinary  materials. For comparison, in an analogous calculation for an ordinary insulator (OI),  we do not find the observed strong effects on bulk plasmons when tuning $\delta$ (see Appendix~\ref{EELS_OI}). Thus, plasmonic control via overlap functions is particularly interesting in models with  non-trivial band structure, which is reflected in the second term of Eq.~\eqref{relative_phase}. Meanwhile, in Fig.~\ref{fig3_eelsSSH}, we notice that the effect vanishes for small momenta $q$, which  matches our observation of the overlap functions in Fig.~\ref{ovpSSH}. For a more quantitative analysis, we perform a small $q$ expansion of the polarization functions (for the complete calculation, see Appendix~\ref{SSH_smallq}). We find that the real space lattice configuration as well as the overlap functions start to affect the intra-(inter-)band polarization function in fourth (second) order of $q$. Therefore, if we evaluate the plasmon frequencies by truncating the polarization functions below these orders, we obtain no $\delta$ dependence in $\omega(q)$, which is depicted as a white dashed line in Fig.~\ref{fig3_eelsSSH}.   

\subsection{Realization of Tuning via Overlap Functions}
\begin{figure}[htbp]
 \includegraphics[width=8.6cm]{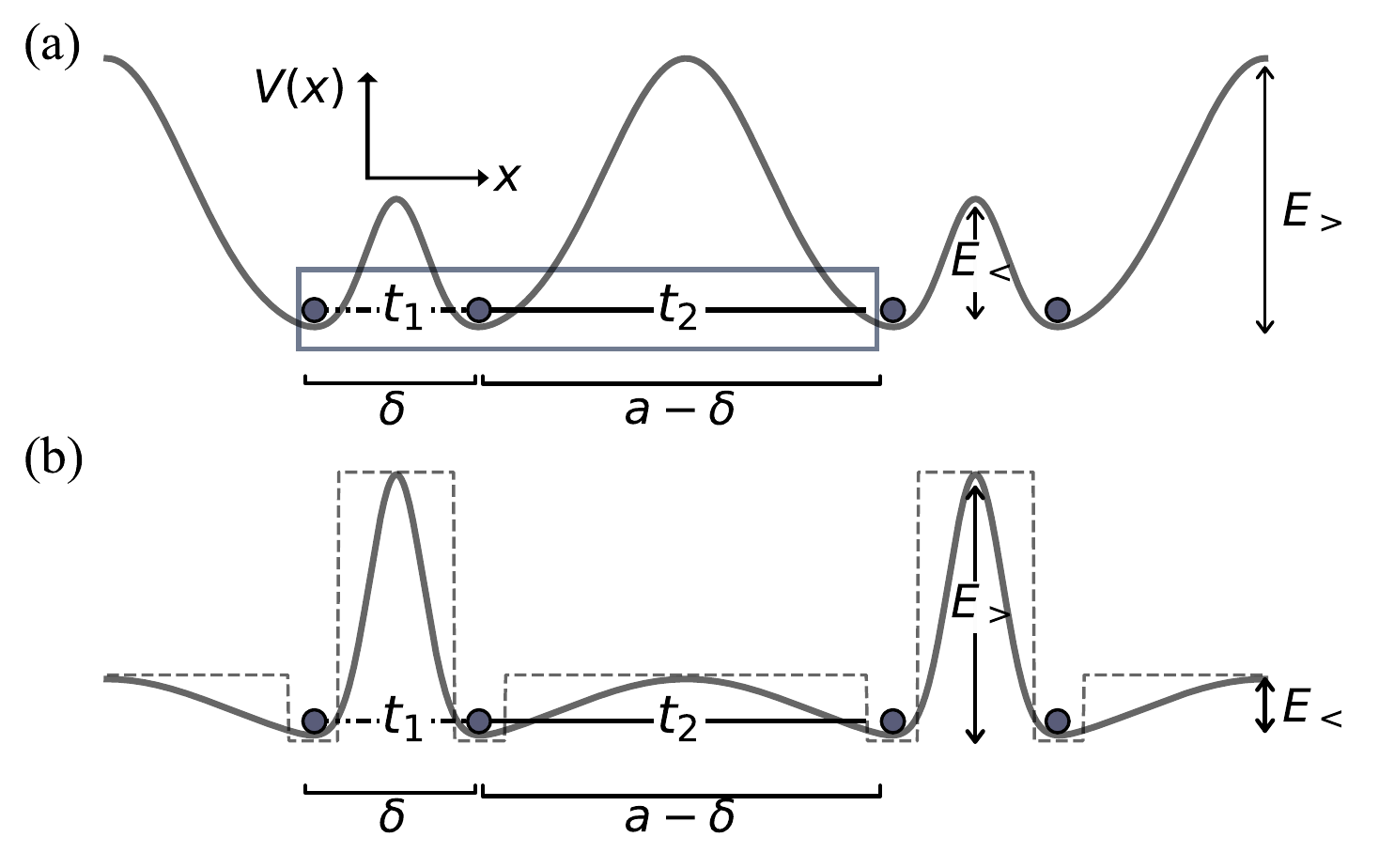} 
 \caption{Illustration of trapped atoms in one-dimensional periodic potential wells. Here,  the separation between basis atoms, $\delta$, is held fixed, and we consider two different regimes: (a) strong intra-basis tunneling, $t_1>t_2$, and (b) strong inter-basis tunneling, $t_2>t_1$.}\label{PotWell}
\end{figure}

 In the above discussion, we considered the separation $\delta$ between two basis atoms A and B as a control parameter that can - in principle - be freely tuned without changing the hopping parameters $t_1$ and $t_2$. In the context of solid materials, this is obviously not realistic, as the hopping matrix element between two sites typically increases when the inter-particle separation is reduced. For example, this is the case when a Peierls instability occurs, whereby the phonon-induced structural dimerization of the underlying lattice leads to an increased hopping matrix element between neighboring atoms within the same basis and a decreased inter-basis hopping. In this case, for example if $t_1=0.5\ \text{eV}$ and $t_2=1.5\ \text{eV}$, it implies $\delta > a-\delta$, i.e., the large-$\delta$ regime appears to be the more natural in the solid state context. For this situation we found above that the low-energy plasmons are dominant, with almost no inter-band screening effect. 
 
However, here we consider an alternative, synthetic, approach to realize this as well as the opposite situation, i.e.,  small inter-particle separation ($\delta < a-\delta$) and small hopping ($t_1<t_2$) at the same time. The system is a one-dimensional array of atoms trapped in periodic potential wells,  illustrated in Fig.~\ref{PotWell}. The solid state case is represented by Fig.~\ref{PotWell}(a), whereas the opposite situation is shown in Fig.~\ref{PotWell}(b). There are two tuning parameters for the potential barrier between two atomic sites: the barrier width and the barrier height. For our purpose, we can associate the potential barrier between two atoms A and B within a unit cell a small width $\delta$ but a large height $E_{>}$. Meanwhile, the potential barrier between atom B and atom A in the next unit cell has a relatively larger width $a-\delta$ ($a$ is the lattice spacing which is always fixed) with a very small height $E_{<}$. For a semi-quantitative estimate, we can approximate these two cases by effective square potential barriers. Then, the tunneling probability through a barrier of width $d$ and height $E_h$ is proportional to $e^{-\sqrt{E_h} d}$. In order to have $t_1<t_2$, we require $e^{-\sqrt{E_>} \delta} < e^{-\sqrt{E_<}(a-\delta)}$, which leads to the condition $\frac{E_>}{E_<} > \frac{(a-\delta)^2}{\delta^2}$.  

\section{\label{sec-4Conclu}Conclusions}
We have analyzed plasmons in a doped topological insulator, i.e. the Su-Schrieffer-Heeger (SSH) chain, with  tunable internal separation $\delta$ between its two basis atomic orbitals within a unit cell. The polarization functions and the dielectric screening effect from the intra-band and the inter-band transitions were investigated analytically, where we found that the overlap functions $O_{nn'}(k,k+q) = |\braket{\Phi_{n,k}|\Phi_{n',k+q}}|^2$ play an important role in determining the collective plasmonic response. Due to the non-trivial topology, the overlap functions of the SSH model show a significant dependence on  $\delta$ for intermediate momenta $q$, which ultimately controls the enhancement or suppression of plasmons excited in different energy regimes. For small momenta, the effects from the overlap functions become negligible because of the vanishing linear term $\propto q$ in the polarization functions, which is a consequence of time-reversal symmetry. Furthermore, we find that such effects are not expected in non-topological doped ordinary insulators. 

We also considered possible experimental realizations. In the solid state setting, where lattice dimerization is naturally accompanied by strong tunneling between neighboring intra-basis atoms, the low-energy plasmons are expected to be dominant in the metallic SSH model, and the high-energy plasmons are weak. In order to experimentally realize the opposite situation, we propose a system consisting of a one-dimensional array of trapped cold atoms, which can enhance the high-energy plasmon branch by tuning the hopping parameters while keeping the basis atom separation fixed.

The effects of overlap functions on plasmons investigated in this work are expected to be general in more complex models and in higher dimensions, especially in cases with non-trivial topology. For example, a simple generalization is the one-dimensional Aubry-André-Harper (AAH) model with three or more basis atoms in a primitive cell. For this model, we find that, when there are three basis atoms, we observe plasmon modes at different energies, which can be tuned by manipulating the separations between the basis atoms. The general complementary property of the overlap functions indicates that we can selectively enhance one plasmon branch and suppress others. However, with further increased number of basis atoms in this 1D model and correspondingly flattened electronic bands, we find that the highest energy plasmon branch is generally becoming  prominent  above the Landau damping regime for a large set of tuning parameters, which is very similar to the recently reported \textit{intrinsically undamped plasmons} in models of twisted-bilayer-graphene \cite{Lewandowski2019}. In this case, other lower-energy plasmon branches which are bounded by the particle-hole continuum show very flat dispersions. These slow plasmons, similar to those reported in generic 2D materials \cite{DaJornada2020}, find promising applications in imaging techniques. And more importantly, they can be sensitively tuned via basis atoms. Recently, we also observed similar effects in  two-dimensional topological insulators \cite{Schlomer2020plasmons}, such as the 2D SSH model \cite{Obana2019}. Besides simply changing the atomic separations in the unit cell, there are  other internal degrees of freedom, such as atomic orbitals of each basis atom, that can be tuned to affect plasmonic or other collective excitations of electrons, which will be  addressed in future work.   

\begin{acknowledgments}
We thank Yuling Guang and Malte R{\"o}sner for useful discussions. This work was supported by the US Department of Energy under grant
number DE-FG03-01ER45908. The numerical computations were carried out on the University of Southern California High Performance Supercomputer Cluster. 
\end{acknowledgments}

\appendix
\section{\label{EELS_OI}Plasmons in an ordinary insulator}
\begin{figure}[htbp]
 \includegraphics[width=8.6cm]{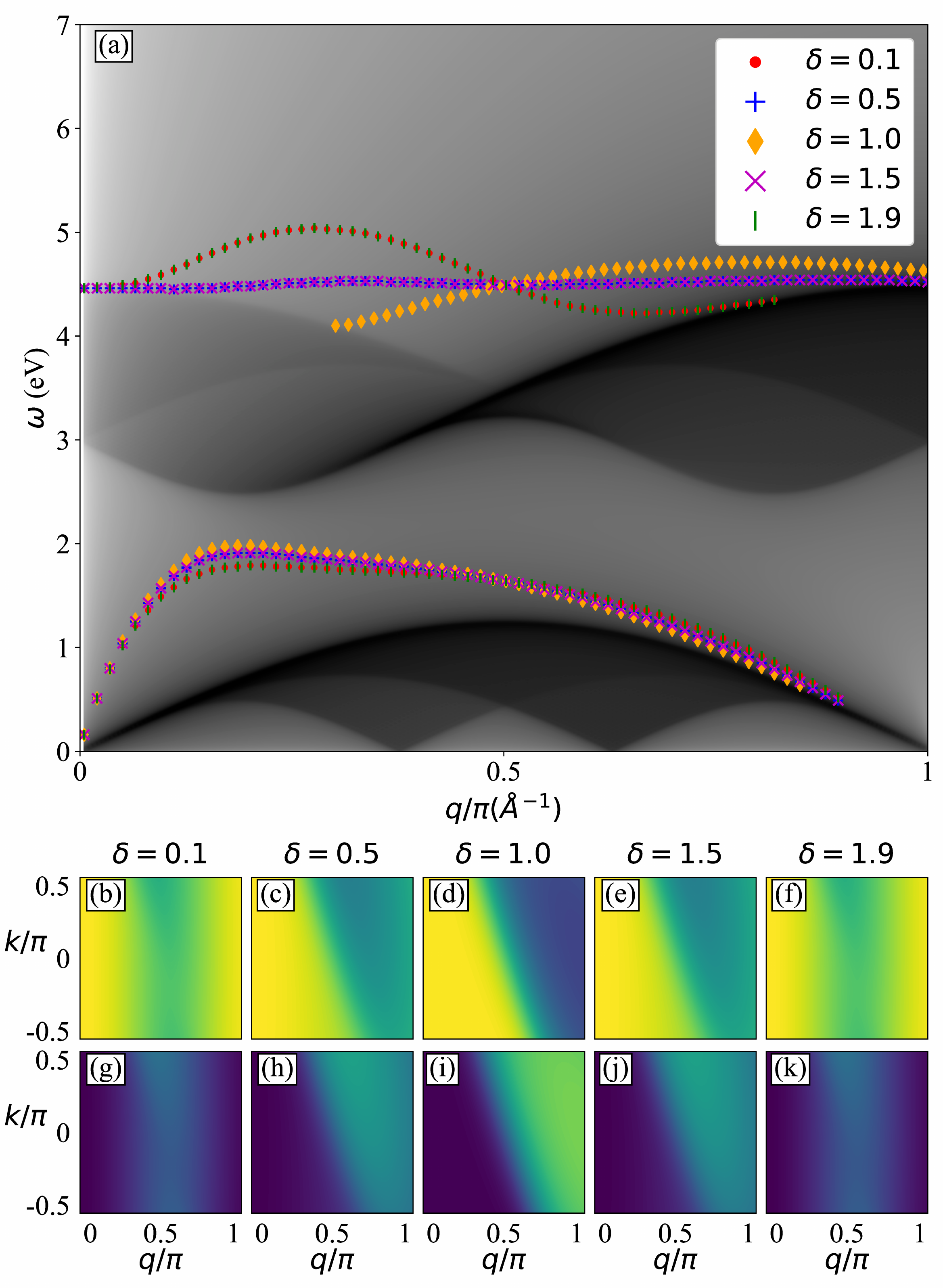} 
 \caption{(a) Plasmon dispersions in an ordinary insulator for different choices of the separation between basis atoms, $\delta$. The dark areas represent the electron-hole continuua. (b)-(k) Overlap functions for the various parameters shown in (a).}\label{figapp_OI}
\end{figure}

As a benchmark, here we determine the plasmonic excitations in an ordinary insulator (OI), parametrized by $t_1=t_2=1\text{ eV}$ and $V_\text{A}=-V_\text{B}=1.0\text{ eV}$. This  parameter set generates the same bandwidth and energy gap as the SSH model discussed in the main text. In the OI, the energy gap is opened up by unequal on-site energies of the two atoms in the unit cell. The plasmon dispersions for different $\delta$ are shown in Fig.~\ref{figapp_OI}(a). Unlike in the SSH model,  no significant tuning effects are observed here. Accordingly, the overlap functions in an OI do not show a clear trend  by tuning $\delta$ [see Figs.~\ref{figapp_OI}(b)-(k)].

\section{\label{SSH_smallq}Random Phase Approximation for plasmons in the SSH model and the ``small-$q$" expansion}
Using the general expression in Eq.~\eqref{Eps_RPA_q} for the dielectric function of two-band models, along with the specific SSH overlap functions in Eqs.~\eqref{ovpSSH} and \eqref{relative_phase}, we can write the polarization function $\chi_0(q,\omega)$ of the SSH model in the form 
\begin{align}
 \chi_0(q,\omega) = 2 \sum_{k,n,n'}f_{n,k}O_{nn'}(k,k+q)T_{nn'}(\omega,k,q), \label{ChiqSimple}   
\end{align}
where
\begin{align}
    T_{nn'}(\omega,k,q) = \frac{2(E_{n',k+q}-E_{n,k})}{(\omega+i0^+)^2 - (E_{n',k+q}-E_{n,k})^2}. \label{Tfunc}
\end{align}

In deriving the above, we used the following properties:\\
(1) change of variables, $k+q \to -k$, justified by the periodicity (in $k$) of the sum in Eq.~\eqref{ChiqSimple}; \\
(2)  time reversal (TR) symmetry, $E_{n,k} = E_{n,-k}$ and $f_{n,k} = f_{n,-k}$; \\
(3) overlap functions obey $O_{nn'}(k,k+q) = O_{n'n}(k+q,k)$, which is easily seen from Eq.~\ref{ovpGeneral}.

In the main text, we consider the SSH model with a partially filled low-energy band $E_{0,k}$, so $f_{1,k}\equiv 0$. Here we perform a small-$q$ expansion for the polarization function. For the intra-band contribution, we set $n=n'=0$, yielding
\begin{align}
 \chi_0^\text{intra}(q,\omega) = 2\sum_{k=-k_F}^{k_F}O_{00}(k,k+q)T_{00}(\omega,k,q).
\end{align}
To obtain the small-$q$ behavior of the intra-band overlap function $O_{00}(k,k+q) = ( 1 + \cos \varphi_{k,q} ) / 2$, we first expand $\cos \varphi_{k,q}$,
\begin{align}
  \cos \varphi_{k,q} &= 1 - \frac{q^2}{2}\left( \partial_q \varphi_{k,q} \right)^2 |_{q=0} \nonumber \\
  &\quad  -\frac{q^3}{2} [\partial_q \varphi_{k,q} \partial_q^2\varphi_{k,q}]|_{q=0} + o(q^4).
\end{align}
There is no linear term in $q$. Thus, for small $q$, the overlap functions have negligible effects on the dispersion, as already discussed in the main text. The cubic term appears because  $\varphi_{k,q} = q\delta + \mathrm{arg}(t_1 + t_2e^{-ika-iqa}) - \mathrm{arg}(t_1 + t_2e^{-ika})$ is neither even nor odd in $q$. The derivatives in the expansion coefficients can be calculated explicitly. They are
\begin{align}
  \partial_q \varphi_{k,q}|_{q=0} &= \delta + \partial_q \mathrm{arg}(t_1 + t_2e^{-ika-iqa})|_{q=0}  \nonumber \\
  &= \delta + a \partial_x \mathrm{arg}(t_1 + t_2e^{-ix}) |_{x=ka} \nonumber \\
  &=\delta - a\frac{t_2^2 + t_1t_2\cos(ka)}{t_1^2 + t_2^2 + 2t_1t_2\cos(ka)} \triangleq \alpha_\delta (k) 
\end{align}
and
\begin{align}
  \partial_q^2 \varphi_{k,q} |_{q=0} &= \partial_k \alpha_\delta (k) = \frac{a^2t_1t_2\sin(ka)(t_1^2 - t_2^2)}{[t_1^2 + t_2^2 + 2t_1t_2\cos(ka)]^2} \triangleq \beta(k).
\end{align}
$\alpha_\delta(k)$ is an even function in $k$ and depends on $\delta$, whereas $\beta(k)$ is an odd function in $k$ and does not depend on $\delta$. Finally, the intra-band overlap function, up to  third order in $q$, is given by
\begin{align}
  O_{00}(k,k+q)  &= 1 - \frac{\alpha_\delta^2(k)q^2}{4} - \frac{\alpha_\delta(k)\beta(k)q^3}{4} + o(q^4). \label{smallq_inta_ovp}
\end{align}
For the small-$q$ expansion of $T_{0,0}(\omega,k,q)$, since we are interested in plasmon solutions where $\omega \gg (E_{n',k+q}-E_{n,k})$, we can write the real part approximately as
\begin{align}
   T_{00}(\omega,k,q) \approx \frac{2}{\omega^2} \bigg( &q\partial_k E_{0,k} + \frac{q^2}{2}\partial_k^2 E_{0,k}  \nonumber \\
   &+ \frac{q^3}{3!}\partial_k^3 E_{0,k} + \frac{q^4}{4!}\partial_k^4 E_{0,k} + \cdots \bigg). \label{smallq_intra_trans}
\end{align}

Combining \eqref{smallq_inta_ovp} and \eqref{smallq_intra_trans}, we can start to look at the small-$q$ behavior of $\chi_0^\text{intra}(q,\omega)$. The summation over $k$ goes from $-k_F$ to $k_F$. The linear term vanishes because $\partial_k E_{0,k}$ is an odd function in $k$. The coefficient (apart from $\frac{4}{\omega^2}$) for the second order expansion, $q^2$, is
\begin{align}
    C_2 = \sum_{k=-k_F}^{k_F} \frac{1}{2} \partial_k^2 E_{0,k}.
\end{align}
For the free electron gas with a quadratic dispersion, this yields the familiar result which is the inverse of the electron mass (apart from a constant factor). This term does not depend on the overlap function and $\delta$. For the $q^3$ term, the coefficient
\begin{align}
    C_3 = \sum_{k=-k_F}^{k_F} \bigg[ \frac{1}{3!}\partial_k^3 E_{0,k} - \frac{1}{4}\alpha_\delta^2(k)\partial_k E_{0,k} \bigg]
\end{align}
also vanishes because the summed term is odd in $k$. The fourth order coefficient is 
\begin{align}
    C_4(\delta) = \sum_{k=-k_F}^{k_F} \bigg[ &\frac{1}{4!}\partial_k^4 E_{0,k} - \frac{1}{8}\alpha_\delta^2(k)\partial_k^2 E_{0,k} \nonumber \\
    &- \frac{1}{4}\alpha_\delta(k) \beta(k) \partial_k E_{0,k} \bigg]. \label{C4}
\end{align}
This is non-vanishing because the summed term is even in $k$, and it implicitly depends on $\delta$ through $\alpha_\delta(k)$. So, for the intra-band polarization function, the overlap function becomes relevant, starting from the $q^4$-term. We stop here and finally write $\chi_0^\text{intra}(q,\omega)$ as
\begin{align}
    \chi_0^\text{intra}(q,\omega) = \frac{4}{\omega^2} [ C_2q^2 + C_4(\delta)q^4 + o(q^5)].
\end{align}

For the inter-band polarization function,  we set $n=0$ and $n'=1$, which gives
\begin{align}
    \chi_0^\text{inter}(q,\omega) = 2\sum_{k=-k_F}^{k_F}O_{01}(k,k+q)T_{01}(\omega,k,q).  \label{ChiqInterSimple}
\end{align}
The small $q$ expansion for $O_{01}(k,k+q)$ can be simply obtained by $1-O_{00}(k,k+q)$, which starts with the $q^2$-term. For $T_{01}(\omega,k,q)$, for simplicity, we use its static limit as an approximation, which turns out to be a good approximation in the low-energy regime. Due to the chiral symmetry of the model, we have $E_{1,k}=-E_{0,k}$. $T_{01}(k,q)$ can then be expanded as
\begin{align}
  T_{01}(k,q) &\approx \frac{1}{E_{0,k}} \bigg[ 1 - \frac{1}{2E_{0,k}}(E_{0,k+q}-E_{0,k}) \nonumber \\
  &\quad + \frac{1}{4E_{0,k}^2}(E_{0,k+q}-E_{0,k})^2 + \cdots \bigg].
\end{align}
Combining $O_{01}(k,k+q)$ and $T_{01}(k,q)$, we observe that there is also no linear term in $q$ in the inter-band polarization $\chi_0^\text{inter}(q)$. The lowest order in $q$ is the quadratic term whose coefficient is
\begin{align}
    \tilde{C}_2(\delta) = \frac{1}{2}\sum_{k=-k_F}^{k_F} \frac{\alpha_\delta^2(k)}{E_{0,k}},
\end{align}
which implicitly depends on $\delta$ via $\alpha_\delta(k)$. Unlike the intra-band polarization, the inter-band polarization already starts to be relevant to the overlap function in order $q^2$. Hence, we write
\begin{align}
  \chi_0^\text{inter}(q) = \tilde{C}_2(\delta)q^2 + o(q^3).  
\end{align}
The inter-band screening is therefore
\begin{align}
    \varepsilon^\text{inter}(q) = 1 - V(q)\tilde{C}_2(\delta)q^2.
\end{align}
Because $\tilde{C}_2$ is negative ($E_{0,k}$ is negative), the inter-band screening is typically larger than 1, which means that the screened Coulomb interaction $W(q)=[\varepsilon^\text{inter}(q)]^{-1}V(q)$ is weakened. Typically, for small $q$, tuning the overlap functions firstly affects the inter-band screening (from $q^2$) and then the intra-band polarization (from $q^4$). The approximated solutions for low-energy plasmons at small $q$ can be determined by setting
\begin{align}
    1 - W(q)\chi_0^\text{intra}(q,\omega_p) = 0.
\end{align}
This gives
\begin{align}
    \omega_p^2 \approx \frac{4V(q)[ C_2q^2 + C_4(\delta)q^4]}{1 - V(q)\tilde{C}_2 (\delta)q^2}. 
\end{align}
The $\delta$-independent part is simply obtained by ignoring $\tilde{C}_2 (\delta)$ and the last two terms in $C_4(\delta)$ [see Eq.~\eqref{C4}]. This yields the white dashed line in Fig.~\ref{fig3_eelsSSH}.

\nocite{*}
\bibliographystyle{apsrev4-2}
\bibliography{bibs}

\end{document}